 \documentstyle{mn}

 \title[Large Scale Inhomogeneity Versus Source Evolution]{
 Large Scale Inhomogeneity Versus Source Evolution \\
 --- Can We Distinguish Them Observationally?
 }

 \author[Nazeem Mustapha et al] 
{Nazeem Mustapha$^{1}$,~
 Charles Hellaby$^{2}$
 ~and G F R Ellis$^{3}$
\\
 {\small Department of Mathematics and Applied Mathematics,} \\
 {\small University of Cape Town, Rondebosch, 7701, South Africa.}\\ 
$^{1}${E-mail:~~{\tt nazeem@gmunu.mth.uct.ac.za}}\\
$^{2}${E-mail:~~{\tt cwh@maths.uct.ac.za}}  \\
$^{3}${E-mail:~~{\tt ellis@maths.uct.ac.za}}
}  

\begin{document}
 \maketitle

 \begin{abstract}
 We reconsider the issue of proving large scale spatial homogeneity of the 
universe, given isotropic observations about us and the possibility of 
source evolution both in numbers and luminosities.  Two theorems make 
precise the freedom available in constructing cosmological models that 
will fit the observations.  They make quite clear that homogeneity cannot 
be proven without either a fully determinate theory of source evolution, 
or availability of distance measures that are independent of source 
evolution.  We contrast this goal with the standard approach that assumes 
spatial homogeneity {\em a priori}, and determines source evolution 
functions on the basis of this assumption. \\[2mm]
 PACS numbers:~~~04.20.-q~,~~~98.50.He~,~~~98.80.-k~,~~~98.80.Dr 
\\
[2mm] 
 Submitted to:~~~Mon. Not. Roy. Astron. Soc.~~~June 1997 
\\
 \end{abstract}

 \section{Introduction}

 Ever since the earliest cosmological models, the Einstein and 
de Sitter models, we have been trying to fit observations to the 
 Friedmann-Lema\^{\i}tre-Robertson-Walker (FLRW) spatially homogeneous 
and isotropic family of models.  The successes of reproducing a Hubble  
 redshift-distance law, calculating the correct cosmic helium \& deuterium  
abundances, and the prediction of a cosmic microwave background radiation 
(CMBR), have convinced us of its validity as a bulk description of the 
universe.  However, {\em proving} that the geometry of the universe is 
FLRW on the largest scales is not easy.  In fact, the history of 
observational cosmology shows that each time improved 
instruments permit deeper surveys, the new data soon reveals 
inhomogeneities on the new scale.  

 The best evidence for homogeneity comes from limits on anisotropy of both 
galaxy counts and the CMBR, obtained in each case by comparison of 
observations in different directions.  However this is strictly speaking 
only evidence 
for isotropy about the earth; homogeneity follows only if we introduce 
a Copernican principle, either for galaxies or for the Cosmic Background 
Radiation \cite{ell75,EGS}.  Without this assumption, the models indicated 
are isotropic about us, but allow a spatial variation of the geometry and 
matter content that is spherically symmetric about our position.%
 \footnote{In a universe that is isotropic but inhomogeneous, there are 
anthropic reasons why we might be near the centre, as argued in 
\cite{EMN78}.}  
 The Copernican principle is not really in dispute on a sub-horizon 
scale,%
 \footnote{Although one can construe the current uncertainties in the 
values of cosmological parameters such as $H_0$ and $\Omega$ as being 
evidence for different values on different scales 
 --- i.e. inhomogeneity, we are not claiming this here.} 
 but could be incorrect on a super-horizon scale if we accept 
theories such as chaotic inflation (\cite{linde}; see also \cite{ell79}).  
Therefore one would like to actually prove homogeneity for the observable 
region of the universe, rather than assuming it on principle, which is 
essentially what happens in the usual approach.  Similar issues have been 
discussed by Goodman \cite{G95}.  

 There are several problems with demonstrating homogeneity from observed 
data.  The deeper observations are not only fainter and redshifted, they 
are also affected by proper motion, reddening and absorption due to 
interstellar matter.  These contribute to selection effects which are 
tricky to compensate for.  But the main problem at large distances is the 
evolution of sources, since deeper observations are received from earlier 
cosmic epochs.  Evolution can take place in source colours, luminosities, 
and sizes; at high redshift it can affect the type of source as well as 
their numbers.  However in this paper, for simplicity, we shall only 
consider bolometric observations of one type of source. 

 How does the number and brightness of the observable sources relate to 
the local density at different times?  Recent evidence for a sharp fall 
off of the space density of quasars above a redshift of $z = 5$ 
\cite{WHO94,SSG95,KDdC95,SWKJH96}, {\em could} be taken as evidence of 
inhomogeneity, though most attribute it to source evolution.  The large 
population of faint blue objects found by sensitive optical surveys is 
thought to be young starforming galaxies at high redshifts, and therefore 
constitute evidence for evolution \cite{JFSEP91,MSFJ91,MSFR95} 
 --- {\em if} we assume the universe has a FLRW geometry. 
Without evolution, these observations are inconsistent with that 
geometry.  One difficulty is that redshifts of faint objects are scarce 
and difficult to obtain.  A redshift of $z \approx 2$ was deduced 
\cite{MSCFG95} by combining number counts versus magnitudes in 3 colours 
with galaxy evolution models and cosmological models, and comparing with 
the few measured redshifts available.  Again this required the assumption 
of homogeneity.%
 \footnote{It turns out many of the sample are not at high redshift, but 
there is still a ``high redshift tail" to the sample \cite{NW95}.)}  
 Indeed, deducing the effects of source evolution by 
comparing observations with predictions in a FLRW model is a standard 
technique.  Similarly studies of a 
 luminosity-size relation also assume a FLRW model 
 --- recent examples are \cite{S93,U95}.  However this cannot lead to 
certainty \cite{ell75}.  The new discovery of a radio galaxy which is 
apparently an immature giant elliptical galaxy at $z = 4.41$ 
\cite{RLBEBG96}, provides a more striking example of probable source 
evolution.  The problem is to show that this is not rather evidence of 
spatial inhomogeneity, manifested in a change of the evolutionary history 
or the nature of objects observed at larger spatial distances from us.  
Also, claims of a periodicity on top of the Hubble law in the 
 redshift-distance relation \cite{BEKS90} indicate significant deviations 
from standard FLRW observational relations, which could be due to spatial 
inhomogeneities (e.g. \cite{EEGMSSTTAF97}) or to a temporal variation in 
the cosmic expansion rate. 

 If suitably smoothed observations are isotropic, the principal 
observations of discrete sources one can hope to make are the number 
counts $n(z)$ of sources as a function of ``distance", conveniently taken 
as given by cosmological redshift $z$, and the magnitudes and angular 
diameters of sources, also as a function of redshift.  If the assumed 
linear size and absolute luminosities of the sources are correct, the 
latter two give the luminosity and area distances $R(z)$, which should be 
equal.  This is rather fortunate since in practice it is often difficult 
to separate the two measures 
 --- one has to define an edge of a galaxy image in order to measure its 
luminosity, and conversely, one often defines the edge relative to the 
central brightness; and both measures are significant in determining 
selection effects \cite{ellper84}.  In any case, at the largest distances 
angular diameters cannot be measured, and it is the luminosity distance 
that is used.  

 We consider two types of source evolution; absolute luminosity $L(z)$, 
and mass per source $m(z)$, i.e. total density over source number density, 
which represents evolution in source number via sources turning on, galaxy 
mergers etc. Since source evolution is likely to be determined as a 
function of age $\tau$, these functions could usefully be expressed as 
$L(\tau(z))$ and $m(\tau(z))$; however it is analytically easier to solve 
the observational equations if they are considered as functions of the 
observable $z$.  This also helps to emphasise that if large-scale 
inhomogeneity were in fact to occur, the age of the universe would vary 
with spatial position and so becomes difficult to handle.%
 \footnote{Since inhomogeneity has been introduced, it is even possible 
the evolution functions are also position dependent.} 

 Earlier work (section 15.3 of \cite{ENSMW85}, section 7 of 
\cite{bi:SEN92}) showed that if the observational relations are isotropic 
and of the FLRW form, then the universe is indeed homogeneous, provided we 
can assume the matter stress tensor is that of pressure-free matter.%
 \footnote{The result should also obtain for barotropic perfect fluids, 
but not necessarily for imperfect fluids.} 
 However that analysis did not fully consider the effects of source 
evolution. 

 In this paper we show that any given isotropic set of observations $n(z)$ \& 
$R(z)$, together with any given evolution functions $L(z)$ and $m(z)$, can 
be fitted by a spherically symmetric dust cosmology 
 --- a Lema\^{\i}tre-Tolman-Bondi (LTB) model --- in which observations 
are spherically symmetric about us because we are located near the central 
 world-line%
 \footnote
 {Despite claims to the contrary in the literature, this is a perfectly 
possible situation \cite{ell75}.}.  
 Thus we show that any spherically symmetric observations we may 
eventually make can be accommodated by appropriate inhomogeneities in a 
LTB model  --- irrespective of what source evolution may occur%
 \footnote{
 This is within the spirit of the programme of observational cosmology set 
out by Kristian and Sachs in their fundamental paper \cite{ks66} and 
developed further by Ellis, Stoeger, et al \cite{ENSMW85}. 
 }. 
Conversely we show that, given any spherically symmetric geometry and any 
set of observations, we can find evolution functions that will make the 
model compatible with the observations. 

 The purpose is to demonstrate explicity
 --- developing the ideas in \cite{ell75} 
 --- that the relationship between the large scale isotropy of 
observations and large scale cosmic homogeneity is weaker than is commonly 
assumed.  Indeed, apart from any other problems, we can't have a good 
demonstration of homogeneity without observational tests of our source 
evolution theories that are independent of cosmological model, or distance 
measures that are not influenced by source evolution.  This emphasises the 
conclusion that if the demonstration of homogeneity depends on knowing the 
source evolution, and validation of source evolution theories depends on 
knowing the cosmological model is homogeneous, then neither is proved.  
Indeed if we do not make the FLRW assumption, our results can be used to 
determine the degree of inhomogeneity from the observations and any given 
source evolution functions.  If we do make the FLRW assumption, they can 
be used to determine the source evolution functions required to make the 
observations compatible with that model.  {\em The latter is the way 
theory is usually run.}  The point of our paper is to emphasise that there 
are other options, and so such source evolution results should be viewed 
with caution. 

 \section{The LTB Model and its Null Cone}

 We here outline the metric and our notation and null cone solution; for 
more details in this notation see \cite{bi:MBHE97}. 

 The general spherically symmetric metric for an irrotational dust matter 
source in synchronous comoving coordinates is the 
 Lema\^{\i}tre-Tolman-Bondi (LTB) \cite{bi:Lem,bi:Tol,bi:Bon} metric 
 \begin{equation} 
   ds^2 = - dt^2 + \frac{(R'(t,r))^2}{1 + 2E(r)} dr^2 + R^2(t,r) \, 
d\Omega^2,      \label{eq:met} 
 \end{equation} 
 where $R'(t,r) = \partial R(t,r)/\partial r$, and $d\Omega^2 = d\theta^2 
+ \sin^2 \, \theta \,d\phi^2$.  The function $R = R(t,r)$ is the areal 
radius, since the proper area of a sphere of coordinate radius $r$ on a 
time slice of constant $t$ is $4 \pi R^2$.  Solving the Einstein field 
equations gives a generalised `Friedmann' equation for $R(t,r)$, 
 \begin{equation} 
   \dot{R}(t,r) = \pm \sqrt{{2 M(r) \over R(t,r)} + 2E(r)} \, ,
        \label{eq:Rdot} 
 \end{equation} 
 and an expression for the density
 \begin{equation}
   4 \pi \rho(t,r) = {M'(r) \over {R^2(t,r) R'(t,r)}}   \label{eq:density} 
 \end{equation} 
 Eq (\ref{eq:Rdot}) can be solved in terms of a parameter $\eta = 
\eta(t,r)$: 
 \begin{equation} R(t,r) = 
   \frac{M(r)}{{\cal E}(r)} \, \phi_0(t,r)  , ~~
     \xi(t,r) = \frac{({\cal E}(r))^{3/2} \,(t - t_B(r))}{M(r)}
        \label{eq:R} 
 \end{equation} 
 where%
 \footnote
 {Strictly speaking, the hyperbolic, parabolic and elliptic solutions 
obtain when $RE/M$ $> 0$, $=0$, \& $<0$ respectively, since $E=0$ at a 
spherical origin in both hyperbolic and elliptic models.}
\[
    {\cal E}(r) = \left\{
      \begin{array}{l}
               2 E(r), \\
               1, \\
               - 2 E(r), 
      \end{array}
             \right.
~~~ \phi_0 = \left\{
      \begin{array}{l}
               \cosh \eta - 1, \\
               (1/2) \eta^2, \\
               1 - \cos \eta, 
      \end{array}
             \right.
\]
\begin{equation}
\xi = \left\{
         \begin{array}{l}
               \sinh \eta - \eta, \\
               (1/6) \eta^3, \\
               \eta - \sin \eta,
         \end{array}
             \right.
   ~~~ \mbox{when~} \left\{
         \begin{array}{l}
               E > 0 \\
               E = 0 \\
               E < 0
         \end{array}
             \right.
       \label{eq:xi_phi0}
 \end{equation}
 for hyperbolic, parabolic and elliptic solutions respectively.

 The LTB model is characterised by 3 arbitrary functions of coordinate 
radius $r$.  $E = E(r) \geq -1/2$ has a geometric role, determining the 
local `embedding angle' of spatial slices, and also a dynamic role, 
determining the local energy per unit mass of the dust particles, and 
hence the type of evolution of $R$.  $M = M(r)$ is the effective 
gravitational mass with comoving radius $r$.  $t_B = t_B(r)$ is the local 
time at which $R = 0$, i.e. the local time of the big bang 
 --- we have a non-simultaneous bang surface.  Specification of these 
three arbitrary functions 
 --- $M(r)$, $E(r)$ and $t_B(r)$
 --- fully determines the model, and whilst all have some type of physical 
or geometric interpretation, they admit a freedom to choose the radial 
coordinate, leaving two physically meaningful choices, e.g. $r = r(M)$, $E 
= E(M)$, $t_B = t_B(M)$. 

 \subsection{The Observer's Null Cone}

 We now take $R$ and $\rho$ as given on the observer's past null cone, and 
we wish to express the 3 arbitrary LTB functions in terms of them, so 
characterising the LTB model that fits the observations.

 We generalise the gauge choice used in \cite{bi:MBHE97} to the case where 
the spatial sections are in general non-flat, i.e. all values of $E$.  
Human observations of the sky are essentially a single event on 
cosmological scales, so we only need to be able to locate a single null 
cone; we don't need a general solution.  On radial null geodesics, $ds^2 = 
0 = d\theta^2 = d\phi^2$; so from (\ref{eq:met}) if the past null cone of 
the observation event $(t = t_0,~r = 0)$ is given by $t = \hat{t}(r)$, 
then $\hat{t}$ satisfies 
 \begin{equation}
   d\hat{t} = - \frac{R'(\hat{t}(r),r)}{\sqrt{1 + 2E}} \, dr 
   = - \frac{\widehat{R'}}{\sqrt{1 + 2E}} \, dr \, .
 \end{equation}
 We will denote a quantity evaluated on the observer's null cone, $t = 
\hat{t}(r)$, by a $\hat{} \,$; for example $R(\hat{t}(r),r) \equiv 
\hat{R}$.  Now if we choose $r$ so that, on the 
past light cone of $(t_0, r)$, 
 \begin{equation}
   \frac{R'(\hat{t}(r),r)}{\sqrt{1 + 2E} \,} = 
   \frac{\widehat{R'}}{\sqrt{1 + 2E} \,} = 1 \, ,    \label{eq:Ref1}
 \end{equation}
 then the incoming radial null geodesics are given by 
 \begin{equation}
   \hat{t}(r) = t_0 - r \, .      \label{eq:ref2}
 \end{equation}

 With our coordinate choice (\ref{eq:Ref1}), the density 
(\ref{eq:density}) and the Friedmann equation (\ref{eq:Rdot}) become 
 \begin{equation} 
   4 \pi \hat{\rho} \hat{R}^2 = {M' \over \sqrt{1 + 2E} \,} 
       \label{eq:nldens} 
 \end{equation} 
 \begin{equation} 
   {\widehat{\dot{R}}} = \pm \sqrt{{2 M \over \hat{R}} + 2E} \, .
     \label{eq:nlrdot} 
 \end{equation} 
 The gauge equation is found from the total derivative of $R$ on the null 
cone
 \begin{equation} 
   \frac{d \hat{R}}{dr} = \widehat{R'} + \widehat{\dot{R}} \, 
     \frac{d \hat{t}}{dr}      \label{eq:Rhat}
 \end{equation} 
 and with (\ref{eq:ref2}) and (\ref{eq:nlrdot}) substituted, it follows 
that 
 \begin{equation} 
   \frac{d \hat{R}}{dr} - \sqrt{1 + 2E} \, = - \widehat{\dot{R}} = 
   - \pm \sqrt{\frac{2M}{\hat{R}} + 2E} \, .    \label{eq:Rhat2} 
 \end{equation} 
 When we solve this for $2E(r)$ by squaring both sides and rearranging, we 
get 
 \begin{equation} 
   1 + 2E = \left\{ \frac{1}{2} \left[ \left( \frac{d \hat{R}}{dr} 
     \right)^2 + 1 \right] - \frac{M}{\hat{R}} \right\}^2 \,\, / \,\,
     \left( \frac{d \hat{R}}{dr} \right)^2 \, .    \label{eq:E} 
 \end{equation} 
 This expression will tell us under what circumstances (or for which 
regions) the spatial sections are hyperbolic $1 + 2E > 1$, parabolic $1 + 
2E = 1$ or elliptic $1 + 2E < 1$, based on data obtained from the null 
cone.  We now use the expression for the density on the null cone to find 
a linear, first order differential equation for $M(r)$.  Eliminating $1 + 
2E$ between (\ref{eq:E}) and (\ref{eq:nldens}), we get 
 \begin{equation} 
   \frac{d M}{dr} + \left({\displaystyle{\frac{4 \pi \hat{\rho} \hat{R}} 
   {\displaystyle{\frac{d \hat{R}}{dr}}}}} \right) \, M = 
   \left( {\displaystyle{ \frac{2 \pi \hat{\rho} \hat{R}^2} 
   {\displaystyle{\frac{d \hat{R}}{dr}}} }} \right) \left[ \left( 
   \frac{d \hat{R}}{dr} \right)^2 + 1 \right] \, .     \label{eq:M} 
 \end{equation} 
 By evaluating (\ref{eq:R}) and (\ref{eq:xi_phi0}) on the null cone we 
find 
 \begin{equation} 
   \hat{R} = \frac{M}{\cal E} \, \hat{\phi_0} \, , ~~~~~~
   \hat{\xi} = \frac{{\cal E}^{3/2} \, \tau}{M}    \label{eq:nlr} 
 \end{equation} 
 where
\[
    {\cal E}(r) = \left\{
      \begin{array}{l}
               2 E(r), \\
               1, \\
               - 2 E(r),
      \end{array}
             \right.
   ~~~ \hat{\phi_0} = \left\{
      \begin{array}{l}
               \cosh \hat{\eta} - 1, \\
               (1/2) \hat{\eta}^2, \\
               1 - \cos \hat{\eta},
      \end{array}
             \right.
\]
\begin{equation}   
 \hat{\xi} = \left\{
         \begin{array}{l}
               \sinh \hat{\eta} - \hat{\eta}, \\
               (1/6) \hat{\eta}^3, \\
               \hat{\eta} - \sin \hat{\eta},
         \end{array}
             \right.
   ~~~ \mbox{when~} \left\{
         \begin{array}{l}
               E > 0 \\
               E = 0 \\
               E < 0
         \end{array}
             \right.
      \label{eq:nlphi}
 \end{equation}
 and 
 \begin{equation} 
   \tau(r) \equiv \hat{t}(r) - t_B(r) = t_0 - r - t_B(r)  \label{eq:tau_r}
 \end{equation} 
 can be interpreted as proper time from the bang surface to the past null 
cone along the particle world lines.  Thus, with $M$ given by (\ref{eq:M}) 
and then $E$ by (\ref{eq:E}), we can solve for $\hat{\eta}$ from 
 \begin{equation} 
   \hat{\phi_0} = \frac{{\cal E} \hat{R}}{M}    \label{eq:phi0_ERM}
 \end{equation} 
 and (\ref{eq:nlphi}), $\tau(r)$ from
 \begin{equation} 
   \tau = \frac{M}{{\cal E}^{3/2}} \, {\hat{\xi}}     \label{eq:tau_ME} 
 \end{equation} 
  with (\ref{eq:nlphi}) again, and hence $t_B(r)$ from (\ref{eq:tau_r}). 

 \subsection{Origin Conditions}
 \label{sec:origin}

 At the origin of spherical coordinates, $r = 0$, where $R(t,0) = 0$ and 
$\dot{R}(t,0) = 0$ for all $t$, we assume that the density is 
 non-zero, that the type of time evolution (hyperbolic, parabolic or 
elliptic) is not different from its immediate neighbourhood, and that all 
functions are smooth
 --- i.e. functions of $r$ have zero first derivative there.  Thus eq 
(\ref{eq:R}) tells us that $RE/M$ and $E^{3/2}/M$ must be finite at $r = 
0$, (\ref{eq:Rdot}) shows us that $E \rightarrow 0$ and hence $M 
\rightarrow 0$ and $E \sim M^{2/3}$ at $r = 0$.  Eqs (\ref{eq:Rhat}) and 
(\ref{eq:Rhat2}) become 
 \begin{equation}
   \left. \widehat{\frac{d {R}}{dr}} \right|_{r=0} = \widehat{R'} |_{r=0} 
   = \left. \frac{d \hat{R}}{dr} \right|_{r=0} = \sqrt{1 + 2E} = 1 \, ,
      \label{eq:Rhat0} 
 \end{equation} 
 and thus $\hat{R} \sim r$ to lowest order near $r = 0$.  From 
(\ref{eq:nldens}) we find 
 \begin{equation}
   M' \approx 4 \pi \hat{\rho_0} r^2 \, , ~~~~~~~~ 
   M \sim \frac{4}{3} \pi \hat{\rho_0} r^3 
 \end{equation}
 and so 
 \begin{equation}
   E \sim \left( \frac{4}{3} \pi \hat{\rho_0} \right)^{2/3} r^2
 \end{equation}
 We verify these origin conditions satisfy (\ref{eq:M}) to order $r^2$ and 
(\ref{eq:E}) trivially to order $r^0$.

 \subsection{Redshift-distance formula}

 We use the fact that in the geometric optics limit, for two light rays 
emitted on the worldline at  $r_{em}$ with time interval $\delta t_{em} = 
t^{+}(r_{em}) - t^{-}(r_{em})$ and observed on the central worldline with 
time interval $\delta t_{ob} =  t^{+}(0) - t^{-}(0)$ 
 \begin{equation} 
   1 + z = \frac{\delta t_{ob}}{\delta t_{em}} \, .    \label{eq:tint} 
 \end{equation} 
 The incoming radial null geodesics are given by 
 \[  dt = - R'(t,r) \, / \, \sqrt{1 + 2E} \, \, dr \, ,  \]
 so for two successive light rays, $-$ \& $+$, passing through two nearby 
comoving worldlines $r_A$ \& $r_B = r_A + dr$ at times $t_A^-$, $t_B^-$, 
$t_A^+$ \& $t_B^+$
 \[
   d (\delta t) = \delta t_B - \delta t_A = dt^+ - dt^- 
   = {\displaystyle{\frac{\left[ - R'(t^{+},r) + R'(t^{-},r) \right]}
     {\sqrt{1 + 2E}}}} 
   \,\, dr
 \]
 Consequently
 \[
   d \ln \delta t   =   - \frac{\partial}{\partial t} \left[ R'(t,r) 
\right] \,/\,\sqrt{1 + 2E} \, \,dr  
 \] 
 which means that, integrating along the light ray and applying this to 
the log of (\ref{eq:tint}), the redshift is given by
 \begin{equation}
   \ln (1 + z) = \int_{0}^{r_{em}} \dot{R}'(t,{r}) \,/\,\sqrt{1 + 2E} 
    \, d{r} \label{eq:lnz}
 \end{equation}
 for the central observer at $r = 0$, receiving signals from an emitter at 
$r = r_{em}$. 

 We need to find the redshift $z$ explicitly in terms of observables.  
 We differentiate (\ref{eq:Rdot}) with respect to $r$: 
 \begin{equation}
   \frac{{\dot{R}}'\,\dot{R}}{\sqrt{1 + 2E} \,} =  
   \frac{M'}{R \sqrt{1 + 2E} \,} - \frac{M R'}{R^2 \sqrt{1 + 2E} \,} + 
   \frac{E'}{\sqrt{1 + 2E} \,} 
 \end{equation}
 so when evaluated on the observer's past null cone, we get
 \begin{equation}
   {\widehat{\frac{{\dot{R}}'}{\sqrt{1 + 2E} \,}}} = 
   \frac{1}{\hat{\dot{R}}} \left[ \frac{M'}{\hat{R} \sqrt{1 + 2E} \,} -  
   \frac{M}{\hat{R}^2} + ({\sqrt{1 + 2E} \,})' \right] 
 \end{equation}
 Now, from (\ref{eq:E}), the derivative of $\sqrt{1 + 2E}$ is given by 
 \begin{equation}
   (\sqrt{1 + 2E} \,)' = \frac{d^2 \hat{R}}{dr^2} - {\displaystyle{\frac{M'}
  {\left( \hat{R} \displaystyle{\frac{d \hat{R}}{dr}}\right)}}} + \frac{M}{\hat{R}^2} - 
   \sqrt{1 + 2E} \, {\displaystyle{\frac{\displaystyle{\frac{d^2
\hat{R}}{dr^2}}}{\displaystyle{\frac{d \hat{R}}{dr}}}}} 
 \end{equation}
 so, after eliminating $M'$ by substituting from equation 
(\ref{eq:nldens}), it follows that 
 \begin{eqnarray}
   {\widehat{\frac{{\dot{R}}'}{\sqrt{1 + 2E} \,}}} 
   &=& \frac{1}{\widehat{\dot{R}}} \left( 4 \pi \hat{\rho} \hat{R} - 
       4 \pi \hat{\rho} \hat{R} \sqrt{1 + 2E} \,\, / \, 
       \frac{d \hat{R}}{dr}
\right.
\nonumber \\
&&
\left.
       + \frac{d^2 \hat{R}}{dr^2} - 
       \frac{d^2 \hat{R}}{dr^2} \sqrt{1 + 2E} \, / \, \frac{d \hat{R}}{dr} 
       \right)   \nonumber \\ 
   &=& - \left( 4 \pi \hat{\rho} \hat{R} + \frac{d^2 \hat{R}}{dr^2} 
       \right) \, / \, \left( \frac{d \hat{R}}{dr} \right) 
 \end{eqnarray}
 where we have used equation (\ref{eq:Rhat2}) to provide the second 
equality.  From (\ref{eq:lnz}) it now follows that 
 \begin{equation}
   \frac{d}{dr} \left[ \ln(1+z) \right] = -\left[ \frac{d^2 \hat{R}}{dr^2} 
   + 4 \pi \hat{\rho} \hat{R} \right] \, / \, \left( \frac{d \hat{R}}{dr} 
   \right) \, ,    \label{eq:Z} 
 \end{equation}
 which theoretically gives the redshift in terms of coordinate radius $r$, 
directly from $\hat{R}(r)$ and $\hat{\rho}(r)$, viz 
 \begin{equation}
   \ln(1 + z) = - \int_0^r \left[ \frac{d^2 \hat{R}}{dr^2} 
   + 4 \pi \hat{\rho} \hat{R} \right] \, / \, \left( \frac{d \hat{R}}{dr} 
   \right) \, dr \,\, .     \label{eq:lnzr_int}
 \end{equation}

 However, we will be given observations in terms of $z$, rather than the 
unobservable coordinate $r$.  This will be addressed in the next section.

 \section{Observables and Source Evolution}

 For simplicity we shall confine ourselves to one type of cosmic source 
and only consider bolometric luminosities.  We shall assume that the 
luminosity of each source can evolve with time, and that the number 
density of sources can also evolve.  The former we write as an absolute 
bolometric luminosity $L$, and the latter we shall represent as an 
evolving mass per source, $m$, which gives the total local density when 
multiplied by the source number density.  As mentioned, we assume isotropy 
about the earth (once our proper motion has been accounted for), and also 
that the post decoupling universe is well described by zero pressure 
matter 
 --- ``dust".  The particles of this dust are galaxies (or perhaps 
clusters of galaxies).  This means we can use the simplest inhomogeneous 
cosmology --- the LTB metric, which is spherically symmetric and 
inhomogeneous in the radial direction only, and is written in comoving 
coordinates.

 The two source evolution functions are most naturally expressed as 
functions of local proper time since the big bang, $L(\tau)$ and 
$m(\tau)$.  However, in a LTB model the time of the bang may vary from 
point to point, so that the age of objects at redshift $z$ is uncertain 
both because the bang time is uncertain and because the location of the 
null cone is uncertain.  
 The proper time from bang to null cone will be a function of redshift, 
$\tau(z)$, and the projections of the evolution functions on the null cone 
we will write as $\hat{L}$ and $\hat{m}$.  
 Of course, $\tau(z)$ is unknown until we have solved for the LTB model 
that fits the data.  However, for the sake of simplicity, we will take 
$\hat{L}$ and $\hat{m}$ to be given as functions of $z$, to illustrate how 
the 3 quantities, cosmic evolution, cosmic spatial variation, and source 
evolution are mixed together in the luminosity and number count 
observations, $\ell$ and $n$.  A treatment dealing with evolution 
functions based on $\tau$ would involve solving a much more complicated 
set of differential equations in parallel. 

 \subsection{Relating Observables to the LTB Model}

 The area distance or equivalently the diameter distance is the true linear 
extent of the source over the measured angular size.  This is by 
definition the same as the areal radius in the LTB model $R$, which 
multiplies the angular displacements to give proper distances 
tangentially.  The projection onto the observer's null cone gives the 
observable quantity $\hat{R}$.  The luminosity distance is theoretically 
the same as the diameter distance \cite{ell71}, and is measurable
provided we know the true absolute luminosity of the source at the time of 
emission $\hat{L}$.  If the observed apparent luminosity is $\ell(z)$ then 
 \begin{equation}
   \hat{R}(z) = 
   \sqrt{\frac{\hat{L}}{\ell}}\,.    \label{eq:R_L}
 \end{equation}

 Let the observed number density of sources in redshift space be $n(z)$ per 
steradian per unit redshift interval, so that the number observed in a 
given redshift interval and solid angle is 
 \begin{equation}
   n d\Omega dz
 \end{equation}
 and over the whole sky this is 
 \begin{equation}
   4 \pi n dz\,.
 \end{equation}
 Thus the total rest mass between $z$ and $z + dz$ is
 \begin{equation}
   4 \pi \hat{m} n dz     \label{mndz}
 \end{equation}
 where $\hat{m}(z) = m(\tau(z))$ is the mass per source 
 --- i.e. the true density over the source number density.  This primarily 
represents the evolution in the number density of sources.  Given a local 
proper density $\rho = \rho(t,r)$, and its value on the null cone 
$\hat{\rho}$, the total rest mass between $r$ and $r + dr$ is 
 \begin{equation}
   \hat{\rho} \widehat{d^3V} = \hat{\rho} \frac{4 \pi \hat{R}^2 
\widehat{R'}}{\sqrt{1 + 2E}} dr     \label{rhodV}
 \end{equation}
 where $\widehat{d^3V}$ is the proper volume on a constant time slice, 
evaluated on the null cone.  Hence by (\ref{mndz}), (\ref{rhodV}) and 
(\ref{eq:Ref1}) 
 \begin{equation}
   \hat{R}^2 \hat{\rho} = \hat{m} n \frac{dz}{dr}\,.     \label{eq:rho_mn}
 \end{equation}
 Thus we may substitute for $\hat{R}$ and $\hat{\rho}$ from (\ref{eq:R_L}) 
and (\ref{eq:rho_mn}). 

 We transform (\ref{eq:Z}) to be in terms of redshift $z$ instead of 
coordinate $r$ by writing it as 
 \[ 
   \frac{d\hat{R}}{dr} \frac{dz}{dr} + \frac{d^2\hat{R}}{dr^2} \, (1+z) 
   + 4 \pi \hat{\rho} \hat{R} (1+z) = 0
 \] 
 and applying 
 \[
   \frac{d \hat{R}}{dz} \frac{dz}{dr} = \frac{d \hat{R}}{dr} \, , ~~~~~~
   \frac{d^2\hat{R}}{dr^2} = \frac{d\hat{R}}{dz} \frac{d^2z}{dr^2} +
   \frac{d^2\hat{R}}{dz^2} \left(\frac{dz}{dr}\right)^2
 \]
 to get 
 \begin{equation}
   \frac{d\hat{R}}{dz} \frac{d^2z}{dr^2}  + \left[ 
   \frac{d^2\hat{R}}{dz^2} \, (1+z) + \frac{\displaystyle{
   \frac{d \hat{R}}{dz}}}{(1 + z)} \right] 
   \left(\frac{dz}{dr}\right)^2 = - 4 \pi \hat{\rho} \hat{R} 
        \label{eq:nraych} 
 \end{equation}
 Integrating with respect to $r$ and using (\ref{eq:rho_mn}) gives 
 \begin{equation}
   \int_{0}^{z} \frac{d}{dr} \left[ \frac{d \overline{z}}{dr} 
       \frac{d\hat{R}}{d\overline{z}} (1 + \overline{z}) \right] dr 
     = -\int_{0}^{z} 4 \pi \hat{\rho}(\overline{z}) 
       \hat{R}(\overline z) (1 + \overline{z}) \frac{dr}{d\overline{z}}
       d \overline{z}
\end{equation}
\begin{equation}
   \frac{dz}{dr} \frac{d\hat{R}}{dz} (1 + z) - 1 
     = -4 \pi \int_{0}^{z} \frac{\hat{m}(\overline{z}) n(\overline{z})}
       {\hat{R}(\overline z)} \, (1 + \overline{z}) \, d \overline{z}
        \label{eq38}
 \end{equation}
 and we used the origin conditions $[(dz/dr)(d\hat{R}/dz)]_0 = 
[d\hat{R}/dr)]_0 = 1$, and $z(0) = 0$.  It follows that 
 \begin{equation}
   \frac{dz}{dr} = \left[ \frac{d \hat{R}}{dz} (1 + z) \right]^{-1}
   \left\{ 1 - 4 \pi \int_{0}^{z} \frac{\hat{m}(\overline{z}) 
   n(\overline{z})} {\hat{R}(\overline z)} \, (1 + \overline{z}) \, 
   d \overline{z} \right\}    \label{eq:fstint}
 \end{equation}
 Note that this equation differs from the analogous one in Stoeger et al 
\cite{bi:SEN92}%
 \footnote 
 {There they use $M_0$ which equals $8 \pi \hat{m} n / \hat{R}^2$ in the 
current notation.} 
 --- their equation (32)
 --- by a factor of $(1+z)$, and perhaps aptly 
illustrates the difference in the coordinate systems.  To get the full 
model we have to solve the null Raychaudhuri equation (\ref{eq:nraych}) to 
get $r(z)$ (and thus $z(r)$).  Equation (\ref{eq:fstint}) is a first 
integral of (\ref{eq:nraych}).  This has to be integrated one more time to 
obtain $r(z)$.  We must also specify boundary conditions at the origin $r 
= 0$, which we have already used in getting to (\ref{eq:fstint}):
 \[
   \frac{dz}{dr}(0) = \frac{dz}{d \hat{R}}(0) \frac{d \hat{R}}{dr}(0)
   = 1 \, / \, \frac{d \hat{R}}{dz}(0)
 \]
 and also
 \[
   z(0) = 0 ~~~~\Leftrightarrow~~~~ r(z = 0) = 0 \, .
 \]
 so that, integrating $dr/dz$ gives
 \begin{eqnarray}
   r(z)& = & \int_0^z \left[ \frac{d \hat{R}}{d\tilde{z}} (1 + \tilde{z}) 
   \right] 
   \nonumber \\
   &\times&
   \left\{ 1 - 4 \pi \int_{0}^{\tilde{z}} \frac{\hat{m}(\overline{z}) 
   n(\overline{z})} {\hat{R}(\overline z)} \, (1 + \overline{z}) \, 
   d \overline{z} \right\}^{-1} \, d\tilde{z} \,\, .     
   \label{eq:rz_int} 
 \end{eqnarray}

 \section{The Theorems}   \label{sec:theorems}

 \subsection{Theorem (A):}~~ Subject to the conditions of appendix 
\ref{ap2}, for any given isotropic observations $\ell(z)$ \& $n(z)$ with 
any given source evolution functions $\hat{L}(z)$ \& $\hat{m}(z)$, a set 
of LTB functions can be found to make the LTB observational relations fit 
the observations. 
 \subsection{Proof: --- Algorithm (A):}~~
 To obtain the {\sc ltb} mass, energy and bangtime functions ($M$, $E$ and 
$t_B$ respectively) from observational data and source evolution we would 
proceed as follows.  
 \begin{itemize}
 \item   Take the discrete observed data for $\ell(z,\theta,\phi)$ and 
$n(z,\theta,\phi)$, average it over all angles to obtain $\ell(z)$ and 
$n(z)$, and fit it to some smooth analytic functions, such as polynomials.  
We may wish to first correct the data for known distortions and selection 
effects due to proper motions, absorption, shot noise, image distortion, 
etc;
 \item   Choose evolution functions $\hat{L}(z)$ and $\hat{m}(z)$ based on 
whatever theoretical arguments may be mustered;
 \item   Determine $\hat{R}(z)$ from $\hat{L}(z)$ and $\ell(z)$ using 
(\ref{eq:R_L});
 \item   Solve (\ref{eq:rz_int}) for $r(z)$ and hence $z(r)$, then convert 
functions of $z$ to functions of $r$ 
 --- see appendix \ref{ap2} for existence conditions;
 \item   Solve (\ref{eq:M}) and (\ref{eq:rho_mn}) for $M(r)$
 --- existence conditions are given in appendix \ref{ap2};
 \item   Determine $E(r)$ from (\ref{eq:E});
 \item   Solve for $\hat{\eta}$ from (\ref{eq:phi0_ERM}) and 
(\ref{eq:nlphi});
 \item   Solve for $\tau(r)$ from (\ref{eq:tau_ME}) and (\ref{eq:nlphi})
 --- $L(\tau)$ and $m(\tau)$ could now be found; 
 \item   Determine $t_B(r)$ from (\ref{eq:tau_r}). 
 \end{itemize}
 In practice, these equations would be solved numerically, and in parallel 
rather than sequentially, nevertheless the above would determine the 
numerical procedure within each integration step. $\Box$

 By determining the 3 arbitrary functions, we have specified the LTB model 
that fits the given observations and evolution functions.  This result
simply asserts we can construct a (generally inhomogeneous) spherically 
symmetric exact solution of the field equations that will fit any 
given source observations combined with 
any chosen source evolution functions. 

 We assert, without proof, that if the given observations and source 
evolution functions are reasonable, then the LTB arbitrary functions will 
generate a reasonable LTB model.  Our definition of `reasonable' is 
intentionally rather vague.  By reasonable observations we obviously 
include the actual data, suitably processed to account for selection 
effects.  We also include `realistic' hypothetical alternatives, but not 
functions that are wildly different from reality.  Reasonable evolution 
functions are hard to define since the actual ones are not well known, 
especially at larger $z$ values.  By a reasonable LTB model, we mostly 
mean that the density and expansion rate will be within realistic ranges.  
A less crucial criterion is that there will be no shell crossings too 
close to the past null cone.  Evolving the model a long time away from the 
null cone, either forwards or backwards, may introduce shell crossings 
because the data is imprecise.  In general we don't expect shell crossings 
on the large scale 
 --- i.e. two or more different large scale flows of galaxies in the same 
region
 --- nevertheless it is conceivable and in that case the LTB description is 
inapplicable.%
 \footnote{Data can be extended through a shell crossing \cite{clarke}, 
but not within the LTB formalism.} 

 \subsection{Corollary (B):}~~ A LTB model can be found to fit the 
observations with zero evolution 
 --- $\hat{m}$ = constant, $\hat{L}$ = constant.
 \subsection{Proof:}~~ This is an obvious consequence of (A).  $\Box$

 Given realistic data, these models will be inhomogeneous.  Indeed this is 
the reason that 
 non-zero evolution functions have been introduced (otherwise, 
observations are incompatible with a FLRW universe). 

 \subsection{Theorem (C):}~~ Subject to the conditions of appendix 
\ref{ap2}, for any given isotropic observations $\ell(z)$ \& $n(z)$, 
and any given LTB model, source evolution functions $\hat{L}(z)$ \& 
$\hat{m}(z)$ can be found that make the LTB observational relations fit 
these observations. 
 \subsection{Proof: --- Algorithm (C):}~~
 We adapt the above algorithmic procedure to prove this.
 \begin{itemize}
 \item   As before, average the data over all angles, and fit it to smooth 
functions $\ell(z)$ and $n(z)$; 
 \item   Specify two of the three functions $M(r)$, $E(r)$ and $t_{B}(r)$, 
the third being determined by the coordinate condition (\ref{eq:Ref1}).  
It seems expedient to choose $M(r)$ and $E(r)$. 
 \item   Determine $\hat{R}(r)$ from the first order differential equation 
in $\hat{R}$ and its $r$ derivative 
 --- equation (\ref{eq:Rhat2}).%
 \footnote
 {Though we don't strictly know the sign of $\widehat{\dot{R}} = 
\sqrt{2M/\hat{R} + 2E}$, it is fairly safe to assume it is positive on our 
past null cone on the large scales we are considering.}
 The functions should be chosen to satisfy the origin conditions of 
section \ref{sec:origin}
 --- see existence conditions in appendix \ref{ap2};
 \item   Calculate $\hat{\rho}(r)$ from (\ref{eq:nldens}); 
 \item   Solve for $t_B(r)$ as well as $\tau(r)$ from (\ref{eq:phi0_ERM}), 
(\ref{eq:tau_ME}) and (\ref{eq:tau_r})) with (\ref{eq:nlphi}) defining 
$\hat{\eta}(r)$; 
 \item   Integrate (\ref{eq:lnzr_int}) to get $z(r)$ 
 --- appendix \ref{ap2} gives the existence conditions;
 \item   Use the given $\ell(z)$ and $n(z)$, to find $\hat{L}(z)$ from 
(\ref{eq:R_L}) and $\hat{m}(z)$ from (\ref{eq:rho_mn}).  From these and 
$\tau(z)$ solve for $L(\tau)$ and $m(\tau)$, if needed.  $\Box$ 
 \end{itemize}

 Again we assert that if the given observations and LTB model are 
`reasonable', then the derived evolution functions will be `reasonable'.  
The idea is that we can vary the LTB model to which we fit the 
observations to some extent, but still keep the required source evolution 
functions within a `realistic' range. 

 \subsection{Corollary (D):}~~ Source evolution functions can be found that 
make the dust FLRW observational relations fit any observations.
 \subsection{Proof:}~~ An obvious consequence of (C).  $\Box$
 \\[3mm]

 \noindent Loosely put these theorems say\\
     (i)~~~You can always fit isotropic observations with an LTB model, 
whatever the source evolution;\\ 
     (ii)~~If you fiddle the source evolution hard enough, you can fit the 
observations to any LTB or dust FLRW model. 

 Although theorem (C) is an extreme case, and is likely to generate highly 
unphysical evolution functions if the LTB model is chosen arbitrarily, it 
is just a generalisation of (D) which is 
regularly used in an attempt to determine evolution functions from 
cosmological observations.  Theorem (C) highlights the dangers of this 
approach.  

 A complication arises if the redshift is not monotonically 
increasing with distance.  We have seen from the well behaved numerical 
example in \cite{bi:MBHE97} for a parabolic case, that $\hat{R}(z)$ and 
$\hat{\rho}(z)$ may not be single valued, and that the 
 $\hat{R}-z$ and 
 $\hat{\rho}$--$z$ plots can loop.  However, in compiling the real 
observational data, we merely add all the galaxies we see at a particular 
redshift, to get a number count.  Similarly, we merely take an average 
over the luminosities observed at a particular redshift, ascribing the 
variation to natural scatter in intrinsic properties and observational 
error, rather than to a multiply valued function.  Thus we make 
$\hat{R}(z)$ and $\hat{\rho}(z)$ single valued by construction.  So the 
data functions we are trying to fit may not lead to such a good 
model.  In other words, assuming we succeed in constructing a well behaved 
LTB model from the data, it may not be the LTB model that best represents 
the real universe.  It seems unlikely 
 --- though not entirely impossible
 --- that there will be a reliable way of 
 de-convolving the superposed parts of these observational data curves, or 
even of discerning whether loops are present.  It is hard to predict how 
likely this scenario is.

 \section{Conclusions}

 We have shown that a LTB model (a Lema\^{\i}tre-Tolman-Bondi spherically 
symmetric dust cosmology) can be found to fit any given set of 
observations of source counts $n(z)$ and luminosity/area distance 
$\hat{R}(z)$, averaged over all angles, and any evolution functions for 
source luminosity $\hat{L}(z)$ and mass per source $\hat{m}(z)$.  In other 
words, even if we accept isotropy, then demonstrating homogeneity 
 --- rather than assuming it must hold because of the Copernican 
principle
 --- requires more than these observations.  Conversely, our result can be 
used to determine the degree of inhomogeneity from the observations and 
given source evolution functions.  

 If the demonstration of homogeneity depends on knowing the source 
evolution, and validation of source evolution theories depends on knowing 
the cosmological model is homogeneous, then neither is proved.  Thus we 
need methods of validating source evolution models that don't depend on 
assumptions of homogeneity to establish the age at any given $z$.  
Similarly deep cosmological distance measures that don't depend on 
luminosity and are not influenced by source evolution would help pin down 
the cosmological model better.  There are various promising developments, 
in particular:

 (a) distance measurement by Supernovae; 

 (b) determinations of cosmological parameters via gravitational 
lensing measurements; 

 (c) accurate measurements of the Sunyaev-Zel'dovich effect;%
 \footnote{Indeed a very interesting option is to use the bounds on the 
inhomogeneity obtained from the Sunyaev-Zel'dovich effect to constrain the 
LTB model chosen, bearing in mind that this method still suffers from 
excessive error due to absorption effects.}

 (d) observation of CMBR Doppler peaks by the MAP and COBRAS/SAMBA%
 \footnote{i.e. Planck Survey}
satellites.  This will only determine parameters in the neighbourhood of 
$z = 1000$, but is independent of source evolution all the same.  

 (e) the increasing number of source evolution studies that look for 
 tell-tale signs of early stages of galaxy evolution, such as intense star 
formation, etc. 

 Once again, the FLRW assumption is usually if not always made in analyses 
of these effects.  A re-analysis that permits inhomogeneity would be very 
worthwhile, as these techniques may well provide information complementary 
to the principal cosmological measures, that would help separate out the 
effects of cosmic evolution, spatial inhomogeneity, and source evolution.  
Some of these issues are discussed in \cite{G95}. 

 In fact, it is already difficult to constrain the values of $H_0$, $q_0$ 
and $\Lambda$ within a homogeneous dust model because of the uncertainty 
in source evolution, as pointed out in \cite{KS93}.  In this case the 
value of $\Lambda$ affects the time evolution of the scale factor, and so 
the deviation of the angular diameter-redshift relation from expectation 
for a $\Lambda=0$ FLRW model could be due to non-zero $\Lambda$ or to 
source evolution.  Similarly the possible presence of 
 non-baryonic dark matter 
 --- or for that matter, the possibility that gravity obeys field 
equations other than Einstein's 
 --- could significantly affect the cosmic time evolution, and 
introduce further uncertainty.

 The introduction of 
 multi-colour observations does not resolve the problem in any simple way.  
If we have observations in various colour bands 
 --- say U B \& V 
 --- then we must replace the source luminosity evolution function by a 
set of evolution functions for the luminosity in each colour.  Thus, if we 
find deviations of the observations from FLRW expectations, we still 
have a freedom to attribute this either to inhomogeneity or to source 
evolution.  It's true that young galaxies with lots of star formation are 
very blue.  But, having introduced colour observations, and permitted 
evolution in colour, we must also admit the possibility of spatial 
inhomogeneities in the intrinsic colours of sources.  We come back to the 
same problem 
 --- are the differences between observations in different colours due to 
source evolution or spatial inhomogeneity?  The only difference here is 
that cosmic evolution is fairly easily factored out, as the redshift is 
measured. 

 We are not here asserting that the observable universe {\em is} 
inhomogeneous, nor are we suggesting that source evolution studies that 
assume homogenity are not worthwhile. 

 The purpose of this paper is to emphasise that we don't have unquestionable
evidence for spatial homogeneity; and that we can't have a good demonstration 
of homogeneity   --- or even homogeneity on average 
 --- without a reliable theory of source evolution, supported by 
measurements that are independent of cosmological model, and/or cosmic 
distance measures that don't depend on knowing the luminosity evolution of 
sources.   Our best basis for assuming spatial homogeneity is the 
 Stoeger-Maartens-Ellis theorem or ``almost EGS theorem" \cite{SME95}, 
which says that, if the universe is expanding and the CMBR (cosmic 
microwave background radiation) is almost isotropic for all observers 
since decoupling, then the universe is almost homogeneous, and more 
specifically, the scale of CMBR anisotropy puts a limit on the degree of 
cosmic inhomogeneity.  But this result depends on a weak form of the 
Copernican principle; and however convincing that principle is in general 
terms, we shouldn't overstate it.  This line of thought says that the 
earth is just another planet around the sun, but it doesn't say all 
planets are the same size or composition.  It says that our galaxy and our 
supercluster are one among many, but allows several types of galaxy and 
considerable variety in galaxy clustering.  Thus the principle does not 
insist on uniformity on any scale, or even that the observable portion of 
the universe has a density particularly close to the ``global average" 
 --- assuming we can define such a thing.  And above all, while it may be 
true in the real universe, it is also possible that this is not so. 

 We are entitled to deduce homogeneity on the basis of untested 
philosophical principles, such as a Copernican principle; but we must be 
quite clear what we are doing when we make such a deduction, and how it 
relates to possible observational tests.  This paper helps throw light on 
the latter issue. \\[3mm]

 \noindent{\Large \bf Acknowledgements}~~~We thank Igor Barashenkov and 
Ronnie Becker for advice on existence of solutions, and Alan Coley and 
Bruce Bassett for helpful comments.  GFRE and CH thank the FRD for 
research grants.

 \appendix

 \section{A Characterisation of Homogeneity from Observations in an 
Isotropic Universe} 

 We demonstrate the procedure in section \ref{sec:theorems} in the case 
where the input observations, after correcting for evolution, are in the 
RW form.  It turns out that the LTB arbitrary functions assume their RW 
form.  This amounts to a proof that a radially inhomogeneous dust universe 
is RW iff the area distance and number count relations as a function of 
redshift take the RW form.  It is a special case of Theorem (B) where we 
assume that there is no evolution.  These RW relations are 
 \begin{equation}
   \hat{R}(z) = {\displaystyle \frac {{q_0}\,{z} + (\,1 - {q_0}\,)\,
 \left( \,1 - \sqrt {2\,{q_0}\,{z} + 1}\,   \right) }{ 
 {H_0}\,{q_0}^{2}\,(\,1 + {z}\,)^{2}}}     \label{eq:RW_Rz}
 \end{equation}
 and 
 \begin{equation}
   \frac{4 \pi \hat{m} n}{3} =  \frac{ \left[ {q_0}\,{z} + 
   (\,1 - {q_0}\,)\, (  \,1 - \sqrt {2\,{q_0}\,{z} + 1}\, ) 
   \right]^{2}}{\,{H_0 q_0^3}(1 + z)^3 \sqrt{2 q_0 z + 1}}     
   \label{eq:RW_mnz}
 \end{equation} 
 respectively.  Then we can integrate the null Raychaudhuri equation
(\ref{eq:nraych}) once obtaining
 \begin{equation}
   \frac{dz}{dr} = H_0 (1 + z)^2 \sqrt{2q_0z + 1}
 \end{equation}
 This may be integrated once again (illustrating with the case $q_0 < {1 
\over 2}$) to obtain  
 \begin{eqnarray} 
   r &=& \frac{1}{H_0 (1 - 2q_0)} \left[ 1 - 
     \frac{\sqrt{2q_0z + 1}}{1 + z} 
\right.
\nonumber \\
&&
\left.   
 + \frac{q_0}{\sqrt{ 1 - 2q_0}} 
     \ln \left( \frac{\sqrt{2q_0z + 1} + \sqrt{ 1 - 2q_0}}
                     {\sqrt{2q_0z + 1} - \sqrt{ 1 - 2q_0}} \right) \right. 
          \nonumber \\ 
     && \left. + \frac{q_0}{\sqrt{ 1 - 2q_0}} \ln \left( 
     \frac{1 - \sqrt{ 1 - 2q_0}}{1 + \sqrt{ 1 - 2q_0}} \right)\right] 
          \label{eq:rz} 
 \end{eqnarray}
 We continue by solving the first order linear differential equation for 
$M(z)$ (the effective gravitational mass) (\ref{eq:M}).  This equation may 
be written as 
 \begin{equation}
   \frac{d}{dz} \left[ \frac{M(z)}{(1 + z) d\hat{R}\,/\,dr} \right] = 
   \frac{2 \pi \hat{m} n}{(1 + z)} +  \frac{2 \pi \hat{m} n}{(1 + z) 
   (d\hat{R}\,/\,dr)^{2}} 
 \end{equation}
 We substitute the RW area distance and number count functions into 
this and find that 
 \begin{equation}
   M(z) = H_0^{2}q_0 \hat{R}^{3} (1 + z)^{3}
 \end{equation}
 and from (\ref{eq:E}) it follows that
 \begin{equation}
   2E(z) = (1 - 2q_0)H_0^{2} \hat{R}^{2} (1 + z)^{2}
 \end{equation}
 These two relations show that $M \propto (2E)^{3/2}$.  We next show that 
this universe has a simultaneous bangtime.  We need (\ref{eq:rz}) in 
addition to (\ref{eq:tau_ME}) and (\ref{eq:nlphi}).  Restricting ourselves 
to the case $q_0 < {1 \over 2}$, it follows that 
 \begin{eqnarray}
   \tau &=& \frac{1}{H_0 (1 - 2q_0)} \left[ \frac{\sqrt{2q_0z + 1}}{1 + z}
                                    \right.
   \nonumber \\
         &&
                                    \left.
- \frac{q_0}{\sqrt{ 1 - 2q_0}} \ln\left(\frac{\sqrt{2q_0z + 1} + \sqrt{ 1 - 
2q_0}}{\sqrt{2q_0z + 1} - \sqrt{ 1 - 2q_0}} \right)
                                    \right] 
 \end{eqnarray}
 and thus 
 \begin{eqnarray}
   t_B(r) &=& t_0 - \frac{1}{H_0 (1 - 2q_0)} 
\left[ 
   1 + 
\right.
\nonumber \\
&&
\left.
   \frac{q_0}{\sqrt{1 - 2q_0}} \ln\left( 
\frac{1 - \sqrt{1 - 2q_0}}{1 + \sqrt{1 - 2q_0}} 
\right)\right] 
 \end{eqnarray}
 which means that the bang surface is simultaneous%
 \footnote
 {We could set $\tau(0)=t_0$ in which case $t_B(r)= 0$, but this is not 
necessary.}. 
 This, together with $M \propto (2E)^{3/2}$ is all we need to show.

 \section{Conditions for Existence of Solutions}
 \label{ap2}

 \subsection{Existence of solutions $r(z)$ and $z(r)$ to equation 
(\protect{\ref{eq:rz_int}})} 

 We know $\hat{R}(z) \rightarrow 0$ as $z \rightarrow 0$, but from 
(\ref{eq:rho_mn}) we expect $n(z) \sim \hat{R}^2(z)$, assuming 
$\hat{\rho}$, $\hat{m}(z)$ and $dz/dr$ all $\sim$ constant as $z 
\rightarrow 0$, and of course $\hat{m}$, $n$, $\hat{R}$ and $(1 + z)$ must 
all be $\geq 0$.  However the existence condition is less stringent. 

 Assume that near $z = 0$, $\hat{m} n (1 + z) / \hat{R} = S(z) z^\sigma$, 
where $\sigma$ is a constant and $S(0)$ is finite and non-zero.  Then, to 
leading order near $z = 0$, 
 \[
   I_1(z) = \int_0^z \frac{\hat{m} n}{\hat{R}} \, (1 + \overline{z}) 
\, d \overline{z}
   = \int_0^z S(\overline{z}) \overline{z}^\sigma dz
 \]
 \[
   I_1 = \left[ S(0) \frac{\overline{z}^{\sigma+1}}{(\sigma + 1)} 
\right]_0^z 
 \]
 which exists provided $\sigma > -1$.  Since all the terms in the 
integrand are positive, $I_1$ is monotonically increasing. 

 We expect $\hat{R}(z)$ to increase from 0 to a maximum, at $z_m$ say, and 
then decrease asymptotically towards 0.  We assume no looping, i.e. 
$\hat{R}(z)$ is single valued and $d\hat{r}/dz$ doesn't diverge, and that 
there is only one maximum.  Thus $d\hat{R}/dz$ goes from positive to 
negative values, and approaches 0 asymptotically.  It is evident from 
(\ref{eq38}) that $4 \pi I_1(z) = 1$ where $d\hat{R}/dz = 0$ in LTB 
models.  Therefore we write $d\hat{R}/dz (1 + z) = P(z) (z - z_m)^\alpha$ 
and $\{ 1 - 4 \pi I_1 \} = Q(z) (z - z_m)^\beta$, where $\alpha$ \& 
$\beta$ are constants, and $P_m = P(z_m)$ \& $Q_m = Q(z_m)$ are finite and 
non-zero.  Then to leading order near $z = z_m$, (\ref{eq:rz_int}) is 
 \[
   r - r_m = \int_{z_m}^z \frac{P(\overline{z})}{Q(\overline{z})} 
(\overline{z} - z_m)^{\alpha - \beta} d\overline{z} 
 \]
 \[
   r = r_m + \left[ \frac{P_m}{Q_m} 
    \frac{(z - z_m)^{\alpha-\beta+1}}{(\alpha - \beta + 1)} 
\right]_{z_m}^z 
 \]
 and so $r(z)$ exists for $\alpha - \beta + 1 > 0$

 Our conditions for the existence of $r(z)$ are:\\
 (i) $\hat{m}$, $n$, $\hat{R}$ and $(1 + z)$ are $\geq 0$, \\
 (ii) near $z = 0$, $\hat{m} n / \hat{R} \sim z^\sigma$ with $\sigma > -
1$, \\ 
 (iii) $d\hat{R}/dz$ is finite everywhere, \\
 (iv) near $z = z_m$, $d\hat{R}/dz (1 + z) \sim (z - z_m)^\alpha$ and $\{ 
1 - 4 \pi I_1 \} \sim (z - z_m)^\beta$, with $\alpha - \beta + 1 > 0$. 

 The condition for the existence of $z(r)$ is \\
 (v) $z(r)$ is monotonic.

 Conditions (i) \& (ii) are manifestly reasonable.  Conditions (iii) \& 
(iv) are more problematic.  As was shown previously \cite{bi:MBHE97}, 
large enough inhomogeneities can create maxima and minima in $z(r)$ and so 
make $r(z)$ multi-valued, especially near $d\hat{R}/dz = 0$, in which case 
neither (iii) nor (v) would be satified.  However, a multi-valued $r(z)$ 
manifested itself in a $\hat{R}(z)$ graph that looped.  In practice, we 
don't expect to get a looping $\hat{R}(z)$ from the observational data.  
The values of $\ell$ and $n$ at each $z$ are averages over all measured 
values, and so are single valued by construction.  Also $z(r)$ was always 
single valued in the numerical examples considered in \cite{bi:MBHE97}, 
so, if $r(z)$ exists, then inverting it should not be a problem.  
Unfortunately (iv) is unlikely to be satisfied {\em exactly} for real data 
 --- the maximum in $\hat{R}$ will not be at exactly the same value as 
the locus of $4 \pi I_1 = 1$, so one would have to tweak the fitted 
function to obtain a numerical solution.  In other words, the function 
$r(z)$ is sensitive to observational error here.  This however is not too 
serious, since there will be a measure of freedom in the smooth functions 
$\ell(z)$ and $n(z)$ that are fitted to the discrete data.  In fact this 
problem exists even if the universe were genuinely homogeneous 
 --- even if we knew the source evolution functions exactly, the best-fit 
curves obtained from imprecise observational data would need adjustment 
to obtain a solution.

 \subsection{Existence of solutions $M(r)$ to equation 
(\protect{\ref{eq:M}})} 

 Eq (\ref{eq:M}) has the form of an inhomogeneous linear first order ODE,
 \[
   \frac{dM}{dr} + a(r) M = b(r)
 \]
 which has the formal solution
 \[
   M = \mu^{-1} \left[ M_m \mu_m + \int_{r_m}^r b \mu dr \right] \, ,~~~~ 
\mu = e^{\int a dr}
 \]
 where $M_m = M(r_m)$ and $\mu_m = \mu(r_m)$.  

 However we know that $d\hat{R}/dr$ goes through 0 at the maximum of 
$\hat{R}(r)$
 --- at $r_m$ say, so both $a(r)$ and $b(r)$ are divergent there.  It is 
evident that $a$ \& $b$ are finite everywhere else, so we just have to 
show $M(r)$ exists in the neighbourhood of this divergence.  Suppose that, 
near $r = r_m$, $d\hat{R}/dr$ is of the form 
 $d\hat{R}/dr \sim (r - r_m)^\nu$, where $\nu > 0$ and is constant, so 
$a(r) = F(r) (r - r_m)^{-\nu}$ and $b(r) = G(r) (r - r_m)^{-\nu}$, where 
$F(r)$ and $G(r)$ are finite, positive and non-zero.  We expect $\nu = 1$.  
Then to leading order near $r_m$, for $\nu = 1$, 
 \[
   \mu = e^{F_m \ln(r - r_m)} = (r - r_m)^{F_m}
 \]
 where $F_m = F(r_m)$, so
 \[
   M = (r - r_m)^{-F_m} \left[ 0 + \int_{r_m}^r G(r) (r - r_m)^{F_m-1} dr 
\right] 
 \]
 \[
   M = (r - r_m)^{-F_m} \left[ \frac{G_m (r - r_m)^{F_m}}{F_m} - 0 \right] 
 \]
 and thus 
 \[
   M = \frac{G_m}{F_m}
 \]
 to leading order.  Comparison with (\ref{eq:M}) shows $M_m = 
\hat{R}_m/2$, which is consistent with the fact that $d\hat{R}/dr = 0$ 
lies on the apparent horizon $R = 2M$.  Thus $M(r)$ exists in the 
neighbourhood of $r_m$.

 For completeness we consider $\nu \neq 1$, working to leading order.
 \[
   \mu = e^{F_m (r - r_m)^{1-\nu}/(1 - \nu)}
 \]
 \begin{eqnarray}
   M &=& e^{-F_m (r - r_m)^{1-\nu}/(1-\nu)}\, \times 
\nonumber \\
&&   
\left[ 
   \int_{r_m}^r G(r) (r - r_m)^{-\nu}  e^{F_m (r - r_m)^{1-\nu}/(1-\nu)} dr 
\right] 
\nonumber
\end{eqnarray}
 \[
   M = e^{-F_m (r - r_m)^{1-\nu}/(1-\nu)} 
   \left[ \frac{G_m e^{F_m (r - r_m)^{1-\nu}/(1-\nu)}}{F_m} 
\right]_{r_m}^r 
 \]
 For $0 < \nu < 1$, we again get
 \[
   M = \frac{G_m}{F_m}
 \]
 to leading order, which is the expected value.  But for $\nu > 1$ we get 
a divergence at $r = r_m$.

 Thus our conditions for existence of $M(r)$ are that \\
 (i) $\hat{m}$, $n$, $\hat{R}$ and $dz/dr$ are $\geq 0$, which ensure 
$\hat{\rho} \geq 0$, and \\
 (ii) $\hat{R}(r) = \hat{R}(z(r))$ has a power-law maximum of the form 
$\hat{R} \sim (r - r_m)^{\alpha}$ with $1 < \alpha \leq 2$, \\
 with a quadratic maximum being the most reasonable. 

 \subsection{Existence of solutions $\hat{R}(r)$ to equation 
(\protect{\ref{eq:Rhat2}})} 

 The equation is
 \[ 
   \frac{d\hat{R}}{dr} = \sqrt{1 + 2E}\, - \sqrt{\frac{2M}{\hat{R}} + 2E} 
 \]
 assuming that we take the positive root on the right
 --- i.e. that large scale recollape has not begun anywhere on our past 
light cone.  Near $r = 0$, $E$, $M$ \& $\hat{R}$ all go to 0, but our 
origin conditions require $M \sim r^3$, $E \sim r^2$ \& $\hat{R} \sim r$, 
so the solution exists here.  Where $\hat{R} = 2M$ the r.h.s. is zero, so 
$\hat{R}$ has a maximum.  We already have $2E \geq -1$ for a well behaved 
metric, and $2M/\hat{R}$ \& $2E$ are separately positive for parabolic and 
hyperbolic models, while for elliptic models we see from (\ref{eq:nlr}) \& 
(\ref{eq:nlphi}) that $(-2E)\hat{R}/M = (1 - \cos \eta) \leq 2$, so 
$\sqrt{2M/\hat{R} + 2E}\, = \sqrt{M/\hat{R}}\,\sqrt{2 + 2E\hat{R}/M}\,$ 
is always real. 

 Our only conditions for $\hat{R}(r)$ to exist are:\\
 (i) the origin conditions of section \ref{sec:origin} are satisfied.

 \subsection{Existence of solutions $z(r)$ to equation 
(\protect{\ref{eq:lnzr_int}})} 

 The origin conditions ensure that, near $r = 0$, $d^2\hat{R}/dr^2 \sim 
0$, $d\hat{R}/dr \sim 1$, and $\hat{\rho} \sim $ constant, so that the 
integral exists in this neighbourhood.

 Where the null cone crosses the apparent horizon, $\hat{R} = 2M$, we have 
$d\hat{R}/dr = 0$.  However, we find from (\ref{eq:Rhat2}) \& 
(\ref{eq:nldens}) that the integrand of (\ref{eq:lnzr_int}) is
 \begin{eqnarray*}
   && \left[ \frac{d^2\hat{R}}{dr^2} + 4 \pi \hat{\rho}\hat{R} \right] / 
   \left( \frac{d\hat{R}}{dr} \right)   \\ \\
   && = 
\left[
   \frac{E'}{\sqrt{1 + 2E}\,} - 
\left( 
   \frac{M'}{\hat{R}} - \frac{M}{\hat{R}^2} 
\left\{
   \sqrt{1 + 2E}\, -  \sqrt{\frac{2M}{\hat{R}}\, + 2E} 
\right\} 
\right.\right.
\nonumber \\
&&
\left. \left.
   + E' 
\right) 
   / \sqrt{\frac{2M}{\hat{R}} + 2E}\, + \frac{M'}{\hat{R} \sqrt{1 + 2E}\,} 
\right] /    
\\ \\
   &&
\left( 
   \sqrt{1 + 2E}\, - \sqrt{\frac{2M}{\hat{R}} + 2E}\, 
\right)    
\\ \\
   && = 
\left[ \left( 
   \frac{M \sqrt{1 + 2E}\,}{\hat{R}^2} - E' - 
   \frac{M'}{\hat{R}} 
       \right) 
\left\{ 
   \sqrt{1 + 2E}\, - \sqrt{\frac{2M}{\hat{R}} + 2E}\, 
\right\}
\right. 
\nonumber \\
&&
\left. 
   / 
\left( 
   \sqrt{1 + 2E}\, \sqrt{\frac{2M}{\hat{R}} + 2E}\, 
\right) 
\right] /    
\left( 
   \sqrt{1 + 2E}\, - \sqrt{\frac{2M}{\hat{R}} + 2E}\, 
\right)    
\\ \\ 
   && = \left[ 
   \frac{M \sqrt{1 + 2E}\,}{\hat{R}^2} - E' - 
   \frac{M'}{\hat{R}} 
        \right] / 
\left( 
   \sqrt{1 + 2E}\, \sqrt{\frac{2M}{\hat{R}} + 2E}\, 
\right) 
 \end{eqnarray*}
 which is well behaved at $\hat{R} = 2M$.

 Our conditions for existence of $\ln(1 + z)$ are merely\\
 (i) the origin conditions, $E \sim r^2$, $M \sim r^3$ near $r = 0$.

 \end{document}